\documentclass[aps,prl,epsfig,floats,twocolumn,superscriptaddress,amssymb,amsmath,showpacs]{revtex4}
\usepackage{graphicx}
\begin{document}
\title{Anomalous Buckling of Charged Rods}

\author{Roya Zandi}
\affiliation{Department of Chemistry and Biochemistry, UCLA, Box
951569, Los Angeles, CA 90095-1569}
\affiliation{Department of Physics, Massachusetts Institute of
Technology, Cambridge, MA 02139, USA}

\author{Ramin Golestanian}
\affiliation{Institute for Advanced Studies in Basic Sciences,
Zanjan 45195-159, Iran}
\affiliation{Institute for Studies in
Theoretical Physics and Mathematics, P.O. Box 19395-5531, Tehran,
Iran}

\author{Joseph Rudnick}
\affiliation{Department of Physics, UCLA, Box 951547, Los Angeles,
CA 90095-1547}

\date{\today}

\begin{abstract}

Unscreened electrostatic interactions exert a profound effect on the
onset of the buckling instability of a charged rod.  When this
interaction is unscreened, the threshold value of the compressional
force needed to induce buckling is independent of rod length for
sufficiently long rods.  In the case of rods of intermediate length,
the critical buckling force crosses over from the classic
inverse-square length dependence to asymptotic length-independent form
with increasing rod length.  It is suggested that this effect might
lead to the possibility of the ``electromechanical'' stiffening of
nanotubes, which would allow relatively long segments of them to be
used as atomic force probes.

\end{abstract}

 \maketitle


When slender objects are subjected to external compressional elastic
forces, they are susceptible to bending deformations.  The onset of
such a deformation is known generically as the buckling instability
\cite{landau,Love}.  Mechanical failure also occurs in elastic and
viscoelastic objects, such as falling ropes \cite{maha1},
``pouring'' filaments of viscoelastic fluids \cite{maha2} as they hit
horizontal solid surfaces, and migrating geophysical fluids reaching
their terminations \cite{geophys}.  An elastic rod of a given material
and length can resist compressional forces up to a so-called {\em
critical buckling force} $F_c$ that increases with the (effective)
bending stiffness of the rod, and decreases with an inverse-square law
with its length $L$ ($F_c \sim 1/L^2$).  Because of this length
dependence, longer filaments possess much lower critical buckling
forces \cite{Love}.

The limitation on rod length imposed by the buckling instability is a
major structural issue in nano-scale mechanics (AFM-tips, nanotubes
\cite{nanotube1} and nanorods \cite{nano2}, etc.)  as well as the
micro- and macro-scale.  However, this limitation results from the
local nature of elasticity, and may in principle be overcome if
long-ranged interactions also exert a stiffening influence.  Here we
study the mechanical response of an elastic charged rod to external
compressional forces, and in particular the onset of Euler buckling
instability, taking into account the nonlocal nature of electrostatic
self-interactions.  For a cylindrical charged rod of radius $r$ and
surface number charge density $\sigma$ , we find that long-ranged
electrostatics leads, in the limit of a long rod, to a non-vanishing
critical buckling force
\begin{equation}
F_c(L \to \infty)=\Delta \; \frac{\pi }{ \varepsilon_0} \; e^2
\sigma^2 r^2,\label{Fc}
\end{equation}
in which $\varepsilon_0$ is the permittivity of free space, $e$ is the
electron charge, and $\Delta=0.1137056$ is a universal numerical
prefactor.  In the case of rods with a finite length $L$, we find that
the above result smoothly crosses over to a local $1/L^2$ dependence
as $L$ decreases.  Crossover to this dependence occurs when $L$ is
small enough that the accumulated  electrostatic self-interaction has
not yet overwhelmed local elasticity.  We also determine the shape of
the rod at the onset of buckling, and show that the buckling rod becomes
considerably flatter in the interior as a result of electrostatic
self-repulsion.

\begin{figure}
\includegraphics[width=.8\columnwidth]{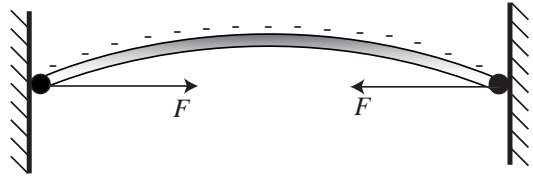}
\caption{The elastic charged rod is hinged at two ends and is
subject to a compressional force $F$.} \label{fig:schematics}
\end{figure}

The elastic charged rod is considered to be inextensible, and its
energy consists of two contributions.  The first part results from the
elastic bending energy $E_{\rm b}=\frac{K}{ 2} \int_0^L d s H(s)^2$.
This energy is controlled by the intrinsic bending modulus $K$ of the
elastic rod and contains no electrostatic contributions.  The
curvature $H(s)$ is assumed to be a function of the arclength
parameter $s$.  For a homogeneous elastic rod of circular cross
section that is made of a material with a Young's modulus $E$, we have
$K=\frac{\pi }{ 4} r^4 E$ \cite{landau}.

The second contribution to the energy arises from electrostatic
interactions, which can be written as $E_{\rm el}=\frac{\Upsilon}{ 2}
\int_0^L d s d s' \frac{1}{|{\bf r}(s) - {\bf r}(s')|}$, for a rod
whose conformation is represented by a space curve ${\bf r}(s)$.  The
electrostatic coupling constant is defined as $\Upsilon=e^2/(4 \pi
\varepsilon_0 a^2)$, $a$ being the average separation between
neighboring charges along the line.  Considering the cylindrical
geometry of the rod, one can express the linear number charge density
$1/a$ in terms of the surface number charge density $\sigma$ through
$a=(2 \pi r \sigma)^{-1}$, which yields $\Upsilon=(\pi/\varepsilon_0)
e^2 \sigma^2 r^2$.

Finally, to study the onset of buckling upon applying a compressive
force $F$, we add a term $E_{\rm ex}=F \int_0^L d s \cos \theta(s)$ to
the total energy.

One can estimate the critical buckling force for a neutral rod using a
simple force balance argument.  Imagine that the rod is bent into an
arc of a circle of radius $R$ with a corresponding arc angle $\theta$,
so that $L=R \theta$.  Then we can calculate the bending energy
$E_{\rm b}(\theta)=\frac{K L}{ 2 R^2}=\frac{K \theta^2}{ 2 L}$, and
the end-to-end distance $x(\theta)=2 R \sin \frac{\theta }{ 2}=L \sin
\frac{\theta }{ 2}/(\frac{\theta }{ 2})$ as a function of the bending
angle $\theta$.  The elastic force that resists bending at the onset
of such arc formation can be found as $F_{\rm b}=-\partial E_{\rm
b}(\theta)/\partial x(\theta)|_{\theta=0}=12 K/L^2$, which slightly
overestimates the exact critical buckling force $F_{c0}=\pi^2 K/L^2$
\cite{landau} due to the artificial assumption of constant curvature.
A similar argument can be used to qualitatively account for Eq.
(\ref{Fc}) in the case of a charged rod with negligible intrinsic
rigidity.  Imagine that charges of unit magnitude are placed along the
rod in a regular pattern at a distance $a$ from each other (see Fig.
\ref{fig:schematics}).  The electrostatic repulsive force that the
first charge experiences can then be calculated as the sum of the
contributions from all the neighbors, namely, $F_{\rm el}(1)=\Upsilon
(1+1/2^2+1/3^2+ \cdots)=\pi^2 \Upsilon/6$.  If the charged rod is to
undergo compressional failure, the external force has to be greater
than this Coulomb repulsion, thus yielding the scaling form for the
critical buckling force as reported in Eq.  (\ref{Fc}).  To obtain the
correct numerical prefactor, however, one should look collective
failure corresponding to the lowest threshold critical force.

A convenient expansion of the total energy can be carried out in terms
of a suitable deformation field.  For the mechanical response that we
would like to consider, it proves sufficient to focus only on planar
deformations, which are conveniently characterized via the angle
$\theta(s)$ that the rod's local unit tangent vector makes with its
unperturbed orientation.  Expanding the total energy up to second
order then yields \cite{odijk}
\begin{eqnarray}
E_{\rm tot}&=&\frac{K}{2} \int_{0}^{L} d s \left[ \frac{d
\theta(s)}{d s}\right]^{2}-\frac{F}{2} \int_{0}^{L} d s \theta(s)^2\nonumber \\
&+&\frac{\Upsilon }{ 2} \int_0^L d s d s' {\cal L}(s,s') \theta(s)
\theta(s'), \label{E1}
\end{eqnarray}
where the electrostatic kernel is given as
\begin{eqnarray}
&&{\cal L}(s,s')=\int_{0}^{L} d s_1 \int_{s_1}^{L} d s_2 \frac{1}{(s_2-s_1)^3}\nonumber \\
&&\times \left\{(s_2-s_1) \left[\Theta(s-s_1)-\Theta(s-s_2)\right]
\delta(s-s')\right.\nonumber \\
&&-\left.\left[\Theta(s-s_1)-\Theta(s-s_2)\right]
\left[\Theta(s'-s_1)-\Theta(s'-s_2)\right]\right\},\nonumber \\
\label{L(s,s')}
\end{eqnarray}
with $\Theta(s)$ the Heaviside step function.  Note that $H(s)=d
\theta(s)/d s$ in this representation.

The total energy is controlled by the spectrum of the following
{\em total energy} bilinear operator
\begin{equation}
{\cal K}(F;s,s')=\left(-K \frac{d^2 }{ d s^2}-F\right)
\delta(s-s')+\Upsilon {\cal L}(s,s').\label{Kfss'}
\end{equation}
While the eigenfunctions of this operator $\psi_{n}(s)$'s are the same
as those for ${\cal K}(0;s,s')$, the eigenvalues have the form of
${\cal E}_{n}-F$, if ${\cal E}_{n}$'s are the eigenvalues of ${\cal
K}(0;s,s')$.  The onset of Euler instability is then found when the
lowest eigenvalue ${\cal E}_0-F$ passes through zero.  For example, if
we switch off electrostatics by setting $\Upsilon=0$, the lowest
eigenvalue will be $K \pi^2/L^2-F$ that goes non-positive at the
critical force $F_{c0}=\pi^2 K/L^2$.  In Ref.  \cite{long}, we have
been able to calculate the exact spectrum of the electrostatic kernel
${\cal L}(s,s')$.  The eigenvalues are found to be \cite{onehalf}
\begin{equation}
\lambda_k=\frac{1}{2}\left[2 \gamma+\psi(\frac{1}{2}+i
k)+\psi(\frac{1}{2}-i k)+\frac{3-4 k^2}{1+4
k^2}\right],\label{eigen-exact}
\end{equation}
where $\gamma=0.577216$ is the Euler constant, and $\psi(x)=d \ln
\Gamma(x)/dx$ is the digamma function. Using this exact result, we
can determine the critical buckling force for an infinitely long
charge rod (or equivalently a charged rod with negligible
intrinsic stiffness) as reported in Eq. (\ref{Fc}) above, in which
the numerical prefactor is calculated as
$\Delta\equiv\lambda_0=\gamma+\psi(1/2)+3/2 \simeq 0.1137056$. The
existence of a non-zero critical force in this limit is a
manifestation of the long-ranged nature of electrostatic
interactions.

\begin{figure}
\includegraphics[width=.7\columnwidth]{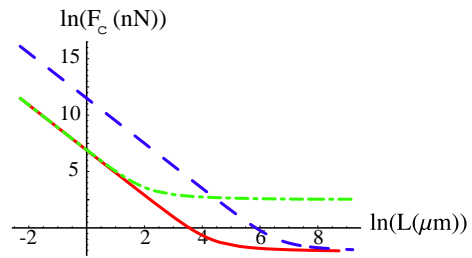}
\caption{The critical buckling force for a charged rod as a function
of its length (log-log plot).  (a) The solid line corresponds to $K=1
\times 10^{-19} \; {\rm N m}^2$ and $\Upsilon=1 \; {\rm n N}$.  (b)
The dashed line corresponds to $K=100 \times 10^{-19} \; {\rm N m}^2$
and $\Upsilon=1 \; {\rm n N}$.  (c) The dash-dotted line corresponds
to $K=1 \times 10^{-19} \; {\rm N m}^2$ and $\Upsilon=100 \; {\rm n
N}$.}
\label{fig:Fc}
\end{figure}

Exactly at the onset of instability, the lowest eigenfunction (the
ground state) $\psi_{0}(s)$ has a vanishing eigenvalue, and thus
provides a nonzero solution to the homogeneous Euler-Largange
equation corresponding to Eq. (\ref{E1}). With the appropriate
boundary conditions, this solution provides the shape of the
elastic charged rod at the onset of instability. We use a cosine
basis expansion for the eigenfunction as $\psi_{0}(s)=\sum_{n=1}
A_n \cos \left(\frac{n \pi s }{ L}\right)$ that corresponds to the
boundary condition of a rod with two hinged ends. The shape of the
deformed rod can be found as
\begin{equation}
u(s)=\int_0^s d s' \psi_{0}(s')=L \sum_{n=1} \frac{A_n}{n \pi}
\sin \left(\frac{n \pi s }{ L}\right).\label{u(s)}
\end{equation}
Note that in the absence of electrostatics we have $u_0(s)=c_0
\sin \left(\frac{\pi s }{ L}\right)$ \cite{landau}.

We study the spectrum of the total energy operator ${\cal K}(0;s,s')$
numerically \cite{details}, and use it to find the critical buckling
force and the shape of a charged rod of arbitrary length at the onset
of compressive failure.  In Fig.  \ref{fig:Fc}, the critical buckling
force is plotted as a function of the length of the rod, for various
values of $K$ and $\Upsilon$.  The plot shows an small-$L$ $1/L^2$
dependence for the critical force that crosses over to an
$L$-independent saturation value as $L$ passes through a crossover
length scale $\ell_{\times}$, where $\ell_\times \propto
\sqrt{K/\Upsilon}$.

\begin{figure}
\includegraphics[width=.7\columnwidth]{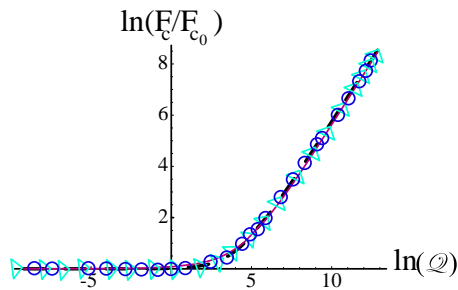}
\caption{The rescaled critical buckling force for a charged rod as
a function of the charging parameter ${\cal Q}=\Upsilon L^2/K$.
Note that three distinct series of data (hollow circles,
triangles, and dashed line), corresponding to the different curves
in Fig. \ref{fig:Fc}, have been collapsed on top of a universal
curve. The solid line represents the interpolation formula of Eq.
(\ref{cal-Q}) for comparison.} \label{fig:universal}
\end{figure}

The critical buckling forces corresponding to different values of
the parameters $K$, $\Upsilon$, and $L$ can be collapsed onto a
universal curve as shown in Fig. \ref{fig:universal}, if
normalized with the critical buckling force of the neutral chain
$F_{c0}$ and plotted as a function of the dimensionless {\em
charging parameter} ${\cal Q}=\Upsilon L^2/K$. An interpolation
formula of the form
\begin{equation}
\frac{F_c }{ F_{c0}}=1+\frac{1 }{ \pi^2} \sqrt{\frac{{\cal Q} }{
2}}+\frac{\Delta }{ \pi^2} {\cal Q},\label{cal-Q}
\end{equation}
is found to satisfactorily represent the universal curve as revealed
by the comparison in Fig.  \ref{fig:universal}.

\begin{figure}
\includegraphics[width=1\columnwidth]{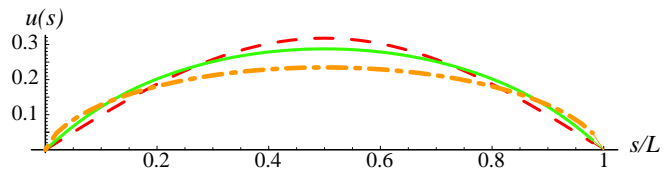}
\caption{The shape of a charged rod at the onset of Euler buckling
instability. The dashed line corresponds to ${\cal Q}=0$, the
solid line corresponds to ${\cal Q}=10^3$, and the dash-dotted
line corresponds to ${\cal Q}=10^6$. The buckling charged rod
flattens in the interior as the charging is increased.}
\label{fig:shape}
\end{figure}

The shape of the charged rod at the onset of buckling is also
calculated, and shown in Fig.  \ref{fig:shape} for various values of
the charging parameter ${\cal Q}$.  It appears that charging leads to
deviations in the shape of the buckling rod from the sinus-profile
\cite{landau} in that there is enhanced flattening in the interior.
This is to be expected because the interior of the charged rod is
subject to stronger build-up of electrostatic self-repulsion as
compared to the end-segments where ``half'' of the repelling charges
are absent.  A similar effect has also been observed in the bending
response of charged elastic rods \cite{bending}.

We can also study the buckling instability for charged rodlike
polymers, or polyelectrolytes, in a solution by using the screened
Debye-H\"uckel interaction $e^{- \kappa r}/(\epsilon r)$ instead
of the long-ranged Coulomb $1/r$ interaction, where $\epsilon$ is
the zero-frequency dielectric constant of the solution and
$\kappa^{-1}$ represents the Debye screening length. The
eigenvalues of the total energy can be calculated numerically for
the screened case, and can be used to determine $F_c$ for each set
of parameters. We find that for sufficiently long polyelectrolytes
the critical buckling force is given as $F_c=\pi^2 k_{\rm B} T
(\ell_0+\ell_{\rm OSF})/L^2$, in which $\ell_0=K/k_{\rm B} T$ is
the intrinsic persistence length, and $\ell_{\rm OSF}=\Upsilon/(4
\epsilon k_{\rm B} T \kappa^2)$ is the well-known
Odijk-Skolnick-Fixman electrostatic persistence length for
polyelectrolytes \cite{odijk,fixman}. This result confirms that
the so-called wormlike chain description of polyelectrolytes
remains valid in determining their mechanical response and onset
of buckling instability, as long as $\kappa L \gg 1$
\cite{long,short}. For shorter chains, however, the $1/L^2$
dependence is altered, and a crossover similar to the one
described in Fig. \ref{fig:Fc} takes place. As a particular case
of interest for assessing the wormlike chain model of
polyelectrolytes one can consider the unscreened limit
($\kappa=0$), where it predicts a crossover of the form
$F_c=\frac{\pi^2 K}{L^2}+\frac{\pi^2}{72} \Upsilon=F_{c0}
\left(1+\frac{0.137078}{\pi^2}{\cal Q}\right)$ \cite{odijk}, which
is to be contrasted with Eq. (\ref{cal-Q}) above.

The familiar image of a long-haired girl touching the van de
Graaff machine suggests that a practical way of imposing the
required charging is by applying a voltage.  For a conducting
cylinder of length $L$ and radius $r$ that is kept at a potential
$V$ relative to ``infinity'' \cite{note}, one can calculate the
induced surface charge density, and deduce from it the
corresponding asymptotic critical buckling force as
\begin{equation}
F_c(L \to \infty,V)=\frac{\Delta \pi \varepsilon_0
V^2}{\left[\ln(L/r)\right]^2}.\label{FcV}
\end{equation}
For a thread of human hair we have $r \simeq 0.1 \;{\rm mm}$ and
$K_{\rm hair} \sim 10^{-11}\; {\rm N m}^2$, which yields for $L=1
\; {\rm cm}$ a critical force of $F_c \simeq 10^{-6} \; {\rm N}$.
Applying a voltage of $V=50 \; {\rm kV}$ (typical of van de Graaff
generators) then results in a critical force of $F_c \simeq
10^{-4} \; {\rm N}$ for a one-meter long piece of hair!

Perhaps the most interesting venue in which these results find
application will be in hardening of atomic force probes.  Carbon
nanotubes have been found to be structurally quite robust and have
exceptionally high Young's modulus (in the TPa range) \cite{young}.
However, the fact that they can grow up to microns in length while
having nanometric diameters renders them quite susceptible to
buckling.  The buckling of multi-walled carbon nanotubes have been
recently investigated experimentally by Dong {\em et al.}
\cite{dong}.  In their experiment, a $6.9 \; \mu {\rm m}$ long
nanotube have been placed under compression and its critical buckling
force (typically in the nN range for micron-sized nanotubes) have been
measured, from which they could extract the bending rigidity of the
nanotube as $K_{\rm nanotube}=8.641 \times 10^{-20} \;{\rm N m}^2$
\cite{dong}.  Using Eq.  (\ref{FcV}), we can now estimate that
a carbon nanotube $30 \;{\rm nm}$ in diameter can resist forces up to
$\sim 1 \; {\rm nN}$ even when it is $1 \;{\rm mm}$ long, provided it
is kept at a voltage of $200 \; {\rm V}$.  This implies a remarkable
``electromechanical stiffness,'' in contrast with the intrinsic
mechanical resistance to buckling which is diminished by a factor of
$10^6$ when the length of the nanotube is increased from a micron to a
millimeter.  The stiffening mechanism may also be useful in the
recently reported nanometer-scale electromechanical actuator that is
based on a multi-walled carbon nanotube \cite{nanoact}.

The authors would like to acknowledge helpful discussions with D.
Chatenay, L. Dong, and T.B. Liverpool.  RZ acknowledges support from
the UC President's Postdoctrol Fellowship program.  This research was
supported by the National Science Foundation under Grant No.
CHE99-88651.


\begin{thebibliography}{99}

\bibitem{Love}
A.E.H. Love, {\it A Treatise on the Mathematical Theory of
Elasticity}, 2nd edition (Dover, London, 1944).

\bibitem{maha1}
L. Mahadevan and J.B. Keller, Proc. R. Soc. Lond. A {\bf 452},
1679 (1996).

\bibitem{maha2}
L. Mahadevan, W.S. Ryu, and A.D. Samuel, Nature {\bf 392}, 140
(1998); Erratum: L. Mahadevan, W.S. Ryu, and A.D. Samuel, Nature
{\bf 403}, 502 (2000).

\bibitem{geophys}
A.M. Johnson and R.C. Fletcher, {\it Folding of Viscous Layers}
(Columbia, New York, 1994).

\bibitem{nanotube1}
B.I. Yakobson, C.J. Brabec, and J. Bernholc, Phys. Rev. Lett. {\bf
76}, 2511 (1996).

\bibitem{nano2}
E.W. Wong, P.E. Sheehan, and C.M. Lieber, Science {\bf 277}, 1971
(1997).

\bibitem{landau}
L.D. Landau and E.M. Lifshitz, {\it Theory of Elasticity}, 3rd
edition (Butterworth-Heinemann, Oxford, 1986).

\bibitem{odijk}
T. Odijk, {J. Polym. Sci.} {\bf 15}, 477 (1977).

\bibitem{long}
R. Zandi, J. Rudnick, and R. Golestanian, Phys. Rev. E {\bf 67},
021803 (2003).

\bibitem{onehalf}
We are correcting here a missing factor of $\frac{1}{2}$ in the
results reported in Ref. \cite{long} [required in Eqs. (B12) and
(B15) therein].

\bibitem{details}
The details of the numerical method is explained in Ref.
\cite{long}.

\bibitem{bending}
R. Zandi, J. Rudnick, and R. Golestanian, Phys. Rev. E {\bf 67},
061805 (2003).

\bibitem{fixman}
J. Skolnick and M. Fixman, {Macromolecules} {\bf 10}, 944 (1977).

\bibitem{short}
R. Zandi, J. Rudnick, and R. Golestanian, Eur. Phys. J. E {\bf 9},
41 (2002).

\bibitem{note}
If a conducting rod is held at a constant electrostatic potential,
the charge distribution along the rod is known to be nearly
uniform. The slight logarithmic buildup of charge at the ends of
the rod leads to no significant change in the electrostatic
energetics of the buckled rod; the conclusions reported in the
body [Eq. (\ref{FcV}) and the related discussions] are not
materially affected.

\bibitem{young}
M.M.J. Treacy, T.W. Ebbesen, and J.M. Gibson, Nature {\bf 381},
678 (1996).

\bibitem{dong}
L. Dong, F. Arai, and T. Fukuda, Proceedings of the 2001 IEEE
Intenrational Conference on Robotics \& Automation (Seoul, Korea,
May 21-26, 2001); J. of Robotics and Mechatronics {\bf 14}, 245
(2002).

\bibitem{nanoact}
A.M. Fennimore, T.D. Yuzvinsky, W.Q. Han, M.S. Fuhrer, J. Cumings,
and A. Zettl, Nature {\bf 424}, 408 (2003).

\end{thebibliography}
\end{document}